\documentclass[conference]{IEEEtran}
\IEEEoverridecommandlockouts

\usepackage{mathrsfs}
\usepackage{xcolor}

\ifCLASSINFOpdf
   \usepackage[pdftex]{graphicx}
 
\else
   \usepackage[dvips]{graphicx}
\fi

\usepackage{amsmath,amssymb}

\usepackage{array}

\usepackage{fixltx2e}

\usepackage{stfloats}

\begin{document}

\makeatletter
\newcommand{\vast}{\bBigg@{4}}
\newcommand{\Vast}{\bBigg@{5}}
\makeatother

\title{Power Module (PM) core-specific parameters for a detailed design-oriented inductor model}

\author{\IEEEauthorblockN{1\textsuperscript{st} Andr\'es Vazquez Sieber}
\IEEEauthorblockA{\textit{* Departamento de Electr\'onica} \\
\textit{Facultad de Ciencias Exactas, Ingenier\'ia y Agrimensura} \\
\textit{Universidad Nacional de Rosario} \\
\textit{** Grupo Simulaci\'on y Control de Sistemas F\'isicos} \\
\textit{CIFASIS-CONICET-UNR} \\
Rosario, Argentina \\
avazquez@fceia.unr.edu.ar}
\and
\IEEEauthorblockN{2\textsuperscript{nd} M\'onica Romero}
\IEEEauthorblockA{\textit{* Departamento de Electr\'onica} \\
\textit{Facultad de Ciencias Exactas, Ingenier\'ia y Agrimensura} \\
\textit{Universidad Nacional de Rosario}\\
\textit{** Grupo Simulaci\'on y Control de Sistemas F\'isicos} \\
\textit{CIFASIS-CONICET-UNR} \\
Rosario, Argentina \\
mromero@fceia.unr.edu.ar}
}

\maketitle

\begin{abstract}
This paper obtains shape-related parameters and functions of a Power Module ferrite core for a design-oriented inductor model, which is a fundamental tool to design any electronic power converter and its control policy. To improve accuracy, some particular modifications have been introduced into the standardized method of obtaining characteristics core areas and lengths. Also, a novel approach is taken to obtain the air gap reluctance as a function of air gap length for that specific core shape. 
\end{abstract}

\section{Introduction}

\IEEEPARstart{F}{e}rrite-core based inductors are commonly found in the LC output filter of voltage source inverters (VSI) \cite{TradeoffStudyHeatSinkOutputFilterVolumeGaNHEMTBasedSingle-PhaseInverter:Castellazzi}, \cite{HybridActivePowerFilterGaNPowerStage5kWSinglePhaseInverter:Manchia}, %\cite{OutputDv/DtFilterDesignCharacterization10kWSiCInverter:Mertens}, 
as energy storage devices in DC/DC converters %\cite{SmallSaturatingInductorsCompactSwitchingPowerSupplies:RinconMora},
 \cite{200COperationDC-DCConverterSiCPowerDevices:Jordan}, \cite{99EfficientThree-phaseBuck-typeSiCMOSFETPFCRectifierMinimizingLifeCycleCostDCdataCenters:Kolar} and as line-input filters in PFC converters \cite{High-VoltageSiC-BasedBoostPFCLEDApplications:Salamero}, \cite{DesignLossAnalysisHighFrequencyPFCconverter:Song}, among many other power conversion applications. Due to the ferrite material properties, these inductors have to deal with relatively high-frequency currents, sometimes being superimposed on relatively large-amplitude low-frequency currents. It is of paramount importance to design these inductors in a way that a minimum inductance value is always ensured which allows the accurate control and the safe operation of the electronic power converter. 
In order to efficiently design these inductors, a method to find the required minimum number of turns $N_{min}$ and the optimum air gap length $g_{opt}$ to obtain a specified inductance at a certain current level is needed. This method has to be based upon an accurate inductor model which needs to be parametrized, among other things, according to the specific core shape, size and the air gap arrangement employed. 

\begin{figure}
    \centering
	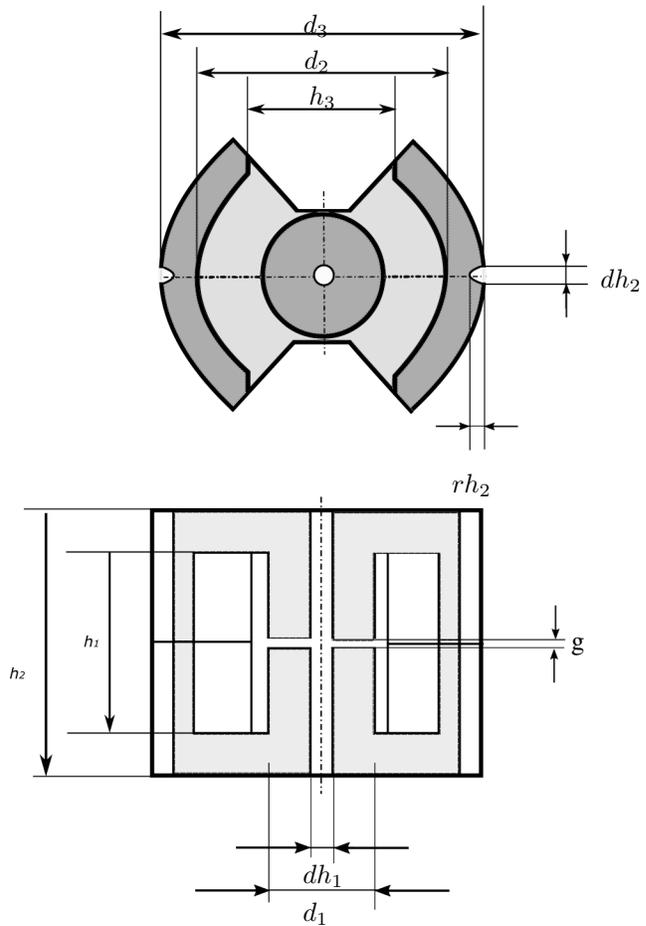
  \caption{Typical PM ferrite core}
  \label{fig:PM_Core_datasheet}
\end{figure}

\begin{figure*}
\centering
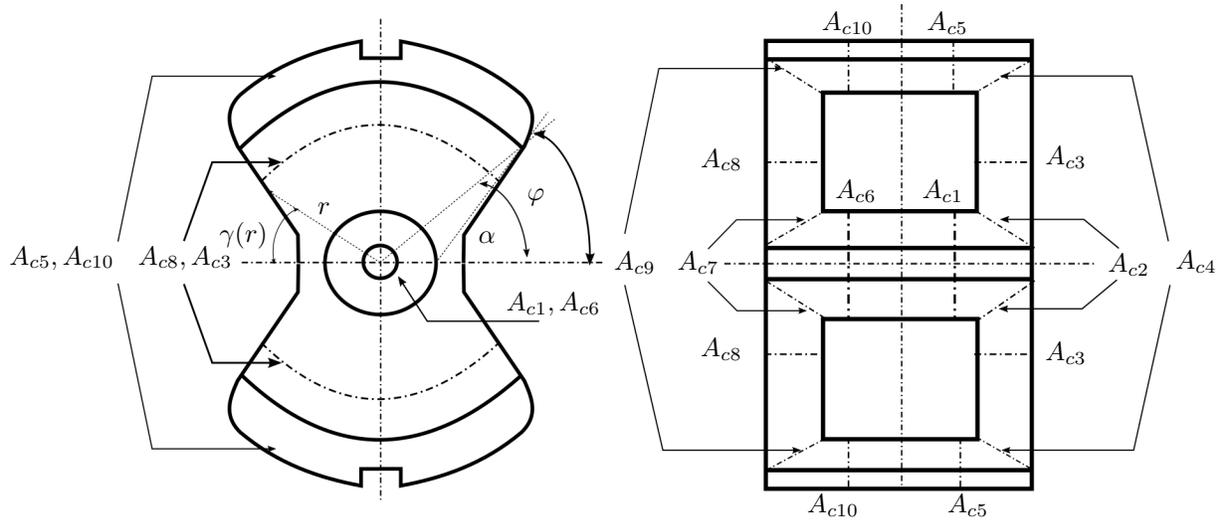
\caption{Cross-sectional areas ($A_{cz}$) corresponding to the zones in which the PM core is divided}
 \label{fig:PM_Core_Ac}
\end{figure*}
It is a standard practice to use the effective core area $A_e$ and length $l_e$ specified in the core datasheet to compute the core reluctance $\mathcal{R}_{c}$, which assumes homogeneous core temperature and magnetic induction \cite{TransformerInductorDesign:McLyman}, \cite{High-FrequencyMagneticComponents:Kazimierczuk}. 
A detailed inductor model would require the determination of a core cross-sectional area vector $\mathcal{A}_c$ and a core length vector $\mathcal{L}_c$, to describe how an ungapped core is divided aiming at a better modelling of $\mathcal{R}_{c}$ and how the ferrite permeability dependence with magnetic induction and temperature impacts on that reluctance. 
Furthermore, the necessary air gap length $g$ to be manufactured has to be obtained as a function of the required air gap reluctance $\mathcal{R}_{gg}$, denoted as $g(\mathcal{R}_{gg})$. This mandates to firstly develop the inverse model, i.e. reluctance as a function of air gap length, $\mathcal{R}_{gg}(g)$, which requires the input of many mechanical dimensions very specific to the ferrite core shape. The usually simplified models frequently require an experimental adjustment of the gap length \cite{PowerElectronics:Mohan}, \cite{InductorsTransformersPowerElectronics:Bossche} and hence a more complex $\mathcal{R}_{gg}(g)$ is needed for a higher accuracy, especially at relatively large air gaps.     
Finally, some dimensions of the corresponding coil former are used to determine the winding utilization factor $k_u$ and the coil DC-resistance $R_{DC}$. 

In this paper, all those parameters and functions are obtained for a specific shape of ferrite core: the Power Module (PM). 
A dimensional drawing of a PM core extracted from a typical datasheet is shown in Figure~\ref{fig:PM_Core_datasheet}. The core depicted has only one air gap placed in the central leg where the two core halves are faced (here referred to as $q_g=1$). Two more gaps similarly located but on each of the external legs can be introduced if a spacer is placed between two equal ungapped ferrite pieces (here referred to as $q_g=2$).
This paper is organized as follows. In section \ref{sec:CoreRegions}, vectors $\mathcal{A}_c$ and $\mathcal{L}_c$ for an ungapped core are obtained. An accurate function $\mathcal{R}_{gg}(g)$ is obtained in section \ref{sec:Reluctance}. A simple method to obtain $g(\mathcal{R}_{gg})$ is presented in section \ref{sec:Length}. Section \ref{sec:CoilFormer} determines the winding height $w_h$, winding width $w_w$ and average turn length $l_t$ based on the dimensions of the coil former. Finally, conclusions are presented in section \ref{sec:Conclusions}. 

\section{Ungapped core regions} \label{sec:CoreRegions}
\begin{figure}
      \centering
			\includegraphics[width=8.5cm]{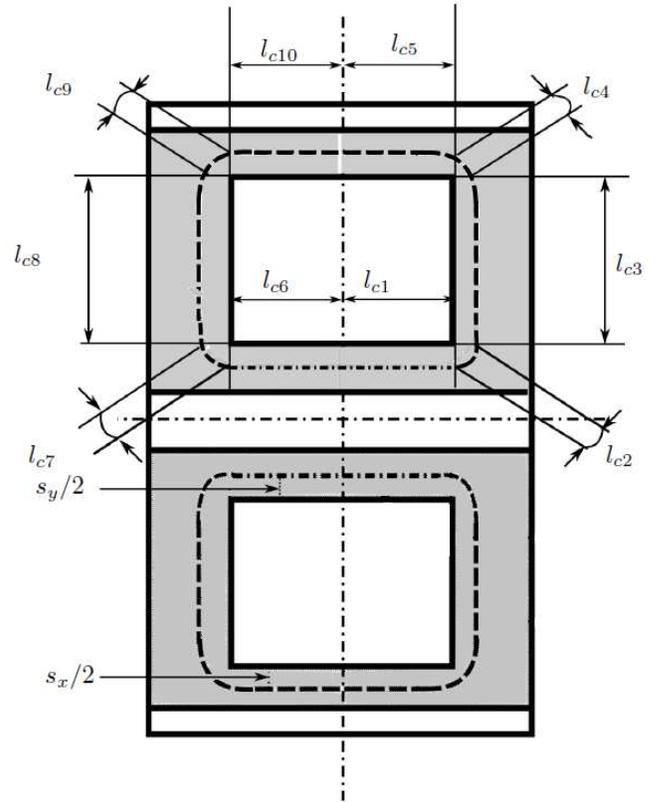}
				\caption{Magnetic path length divisions ($l_{cz}$) corresponding to the zones in which the PM core is divided}
       \label{fig:PM_Core_lc}
\end{figure}

A comprehensive inductor model should divide the ferrite core into a number of $Z$ parts, where $\mathcal{A}_c=[A_{c1} \; ... \; A_{cz} \; ... \; A_{cZ}]$ and $\mathcal{L}_c=[l_{c1} \; ...\; l_{cz} \; ... \; l_{cZ}]$. The PM core is then sectorized in $Z=10$ regions in full compliance with \cite{CalculationEffectiveParametersMagneticPieceParts:IEC60205} and so $z = 1~...~10$. The location of the magnetic path $L_c$, cross-sectional areas $A_{cz}$ and core lengths $l_{cz}$, which are shown in Figure \ref{fig:PM_Core_Ac} and Figure \ref{fig:PM_Core_lc}, are also determined according to \cite{CalculationEffectiveParametersMagneticPieceParts:IEC60205}, with the exceptions that are detailed next. All the needed dimensions to apply \cite{CalculationEffectiveParametersMagneticPieceParts:IEC60205} come from Figure~\ref{fig:PM_Core_datasheet} supposing an ungapped core, i.e. $g=0$, using the average values detailed in the corresponding core datasheet, like for example \cite{FerritePM62/49:TDK-EPCOS}.

We have observed in several numerical examples that $A_{c3}=A_{c8}$ calculated following \cite{CalculationEffectiveParametersMagneticPieceParts:IEC60205} yield smaller values than the correspondingly obtained for $A_{c1}$. This seems to be in contradiction with \cite{PM-cores_made_of_magnetic_oxides_and_associated_parts:IEC61247} which states that the minimum cross-sectional area ($A_{min}$) should coincide with $A_{c1}$ in all PM cores models. To save this discrepancy, we propose firstly that $l_{c3}$ and $l_{c8}$ 
are equal to the distance between the central and the external legs, shown in Figure \ref{fig:PM_Core_lc}. Secondly, $A_{c3}$ and $A_{c8}$ are then adjusted in such a way that

\begin{align}
\label{eq:AreaAdjustment}
\frac{l_{c3}}{A_{c3}}=\frac{l_{c8}}{A_{c8}}=\frac{1}{h_2-h_1} \int_{\frac{d_1}{2}}^{\frac{d_2}{2}} \frac{dr}{r \gamma(r)} 
\end{align}
where $r$ and $\gamma(r)$ are shown in Figure \ref{fig:PM_Core_Ac} for regions $z=3$ and $z=8$. The definite integral in \eqref{eq:AreaAdjustment} %defining $A_{c3}$ and $A_{c8}$ 
has to be numerically solved. The values of $A_{c3}=A_{c8}$ in this way obtained comply with \cite{PM-cores_made_of_magnetic_oxides_and_associated_parts:IEC61247}. 
Other slight divergence exists in the calculation of $l_{c4}$ and $l_{c9}$. This takes into account the alteration of the magnetic path situation along the external legs, due to the notches in $A_{c5}$ and $A_{c10}$, introducing the correction factor $s_x$, which is neglected in \cite{CalculationEffectiveParametersMagneticPieceParts:IEC60205}.

Finally, the equations for each $l_{cz}$ and $A_{cz}$ are 
\begin{itemize}

		\item $l_{c1}=l_{c6}=\frac{h_1}{2}$	
		\item $l_{c2}=l_{c7}=\frac{\pi}{16}(h_2-h_1+2~s_y)$		
		\item $l_{c3}=l_{c8}=\frac{d_2-d_1}{2}$		
		\item $l_{c4}=l_{c9}=\frac{\pi}{16}(h_2-h_1+2~s_x)$
		\item $l_{c5}=l_{c10}=\frac{h_1}{2}$
		\item $A_{c1}=A_{c6}=\frac{\pi}{4}(d_1^2-d_{h1}^2)$
		\item $A_{c2}=A_{c7}=\frac{\pi}{4}\left[\frac{d_1^2-d_{h1}^2}{2}+d_1(h_2-h_1)\right]$
		\item $A_{c3}=A_{c8}=\frac{(h_2-h_1)(d_2-d_1)}{2\int_{\frac{d_1}{2}}^{\frac{d_2}{2}}{\frac{dr}{r\gamma(r)}}}$
		\item $A_{c4}=A_{c9}=(d_3^2-d_2^2)\frac{\frac{\pi}{2}-\alpha}{4}-\frac{\pi}{4}d_{h2}r_{h2}+\frac{\frac{\pi}{2}-\alpha}{2}d_2(h_2-h_1)$ 
		\item $A_{c5}=A_{c10}=(d_3^2-d_2^2)\frac{\frac{\pi}{2}-\alpha}{2}-\frac{\pi}{2}d_{h2}r_{h2}$
	\end{itemize}
where the correction factors are
\begin{align*}
s_y&=d_1-\sqrt{\frac{d_1^2+d_{h1}^2}{2}}\\
s_x&=\sqrt{\frac{d_3^2+d_2^2}{2}-\frac{2\pi}{\frac{\pi}{2}-\alpha}d_{h2}r_{h2}}-d_2\\
\end{align*}
and the required angular variables are 
\begin{align*}
\alpha&=\mathrm{arcsin}\left(\frac{h_3}{d_2}\right)	\\
\gamma(r)&=\pi-2~\mathrm{arcsin} \vast\{ \frac{\mathrm{sin}(\pi-\varphi)}{r} \vast[ \frac{d_1}{2}\mathrm{cos}(\pi-\varphi) \\
 & +\sqrt{r^2-\frac{{d_1}^2}{4}\left[ 1-\mathrm{cos}^2(\pi-\varphi) \right] } \vast]  \vast\} 	\\
\varphi&=\mathrm{arctan} \left(\frac{h_3}{d_2 \mathrm{cos} (\alpha)-d_1}\right)	
\end{align*}

Note that regions located at same relative places in their respective core pieces, actually have same $A_{cz}$ and $l_{cz}$. In fact, \cite{CalculationEffectiveParametersMagneticPieceParts:IEC60205} considers the core divided into only five zones, grouping the pairs with same $A_{cz}$ into a single area with its length doubled, to calculate core effective area $A_e$ and effective length $l_e$. However, taking $Z=10$ does not add an extra calculation effort and allows to consider a different core temperature in each section with equal cross-sectional area, by using a core temperature vector $\mathcal{T}_c=[T_{c1} \; ... \; T_{cz} \; ... \; T_{cZ}]$. 

\section{Air gap reluctance $\mathcal{R}_{gg}$} \label{sec:Reluctance}
			\begin{figure}
       \centering
			\includegraphics[width=8.5cm]{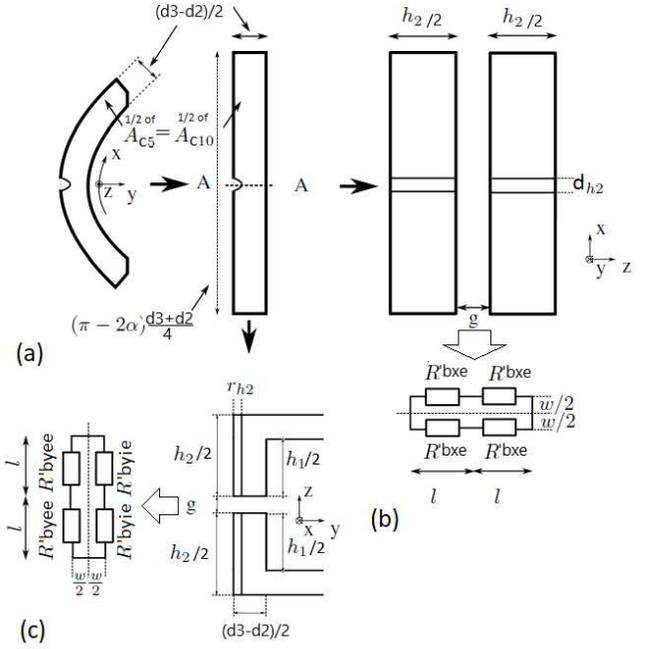}
			\caption{Half external gap in PM core. For $\mathcal{R}^{'}_{bxe}$: $l=\frac{g}{2}$, $w=(\pi-2\alpha)\frac{d_3+d_2}{4}$ and $h=\frac{h_2}{2}$. For $\mathcal{R}^{'}_{byie}$: $l=\frac{g}{2}$, $w=w(x)$ and $h=\frac{h_1}{2}$. For $\mathcal{R}^{'}_{byee}$: $l=\frac{g}{2}$, $w=w(x)$ and $h=\frac{h_2}{2}$.}
       \label{fig:PM_Core_extgap}
\end{figure}

	In a gapped PM core having $q_g=2$, $\mathcal{R}_{gg}$ is composed by the air gap reluctance placed on the central leg, $\mathcal{R}_{ggc}$ plus the combined air gap reluctances located in the external legs, $\mathcal{R}_{gge}$. Assuming an equal distribution of the magnetic flux in both external legs
	\begin{align}
	\label{eq:R_gg_core}
	\mathcal{R}_{gg}&=\mathcal{R}_{ggc}+\mathcal{R}_{gge}
	\end{align}	
If $q_g=1$, then $\mathcal{R}_{gge}=0$. 

\subsection{External reluctance $\mathcal{R}_{gge}$}

\begin{figure}
       \centering
			\includegraphics[width=8.5cm]{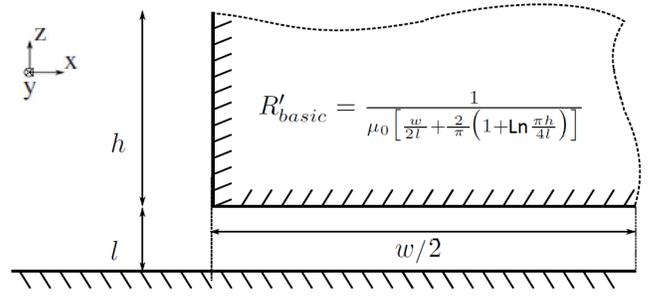}
       \caption{$\mathcal{R}^{'}_{basic}$}
       \label{fig:Rbasic}
\end{figure}

The fundamental idea used to determine $\mathcal{R}_{gge}$, shown in Figure~\ref{fig:PM_Core_extgap}a, is to transform each half of the curved surfaces $A_{c5}$ and $A_{c10}$ into rectangular ones having	the same cross-sectional area, maintaining the same gap length $g$. In this way, the 3D reluctance of this equivalent air gap is addressed using the approach of \cite{3DAirgapReluctanceCalculations:Muhlethaler}. The computation starts considering a basic air gap disposition where the 2D reluctance per unit of length is known. This basic 2D reluctance is referred to as $\mathcal{R}^{'}_{basic}$,
	\begin{align}
	\mathcal{R}^{'}_{basic}&=\frac{1}{\mu_o\left[\frac{w}{2l}+ \frac{2}{\pi}\left(1 + \ln{\frac{\pi h}{4l}}\right)\right]} \notag
	\end{align}			
and is depicted in Figure~\ref{fig:Rbasic}.  $\mathcal{R}^{'}_{basic}$ considers the fringing effects in the magnetic flux near the air gap in the \emph{x-z} plane but it assumes an infinite length in the \emph{y} dimension, i.e. there is no fringing effects in this direction. For the core dimensions shown in Figure~\ref{fig:PM_Core_extgap}b (\emph{x-z} plane), the actual air gap is decomposed into a parallel/series structure comprised of basic gap dispositions each having a reluctance $\mathcal{R}^{'}_{bxe}$ to obtain the 2D reluctance per unit of length,  considering no fringing effects towards the center of the core, i.e. in the \emph{y} dimension. This total reluctance obtained is referred to as  $\mathcal{R}^{'}_{xe}$. Next, the fringing factor $\sigma_{xe}$ is obtained as the relationship between $\mathcal{R}^{'}_{xe}$ and the 2D reluctance per unit of length neglecting the fringing effects.   
			Accordingly, considering the fringing effects in the \emph{x}-direction, the basic 2D reluctance per-unit-of-length $\mathcal{R}^{'}_{bxe}$, the total 2D reluctance per-unit-of-length in the \emph{x-z} plane $\mathcal{R}^{'}_{xe}$ and the fringing factor $\sigma_{xe}$ would be
			\begin{align*}
		\mathcal{R}^{'}_{bxe}&=\frac{1}{\mu_o\left[\frac{(\pi-2\alpha)(d_2+d_3)}{4g}+ \frac{2}{\pi}\left(1 + \ln{\frac{\pi h_2}{4g}}\right)\right]} \\
		\mathcal{R}^{'}_{xe}&=\frac{2 \mathcal{R}^{'}_{bxe}}{2}=\mathcal{R}^{'}_{bxe} \\
		\sigma_{xe}&=\frac{\mathcal{R}^{'}_{xe}}{\frac{4g}{\mu_o(\pi-2\alpha)(d_2+d_3)}} 		
	\end{align*}
			
			For the y-direction, the \emph{y-z} plane shown in Figure~\ref{fig:PM_Core_extgap}c is considered, noting that now there is no fringing flux in the x-dimension. Here there are two basic reluctances $\mathcal{R}^{'}_{byee}$ and $\mathcal{R}^{'}_{byie}$ since the distance from the gap to the next corner of the core in the external side is different from the internal side of the core. Moreover, due to the small notch of radio $r_{h2}$ on the outer side of the core, 
the air gap has no uniform width. In this situation, the gap width $w(x)$, the inner and outer basic reluctances per unit of length, $\mathcal{R}^{'}_{byie}(x)$ and $\mathcal{R}^{'}_{byee}(x)$ respectively, are used to find the fringing factor $\sigma_{ye}(x)$ all along the gap length. Accordingly,
		\begin{align*}
			\mathcal{R}^{'}_{byee}(x)&=\frac{1}{\mu_o\left[\frac{w(x)}{g}+ \frac{2}{\pi}\left(1 + \ln{\frac{\pi h_2}{4g}} \right)\right]} \\
			\mathcal{R}^{'}_{byie}(x)&=\frac{1}{\mu_o\left[\frac{w(x)}{g}+ \frac{2}{\pi}\left(1 + \ln{\frac{\pi h_1}{4g}} \right)\right]}	
			\end{align*}
			where
	\begin{align*}
		w(x)&=\begin{Bmatrix} \frac{d_3-d_2}{2}  & \mbox{if}& a>\left|x\right| >b \\ \frac{d_3-d_2}{2}-r_{h2}\sqrt{1-4\left(\frac{x}{d_{h2}}\right)^2}  & \mbox{if}& b \geq \left|x\right| 
		\end{Bmatrix} \\
		a&=\frac{(\pi-2\alpha)(d_2+d_3)}{8}  \quad \quad b=\frac{d_{h2}}{2}
			\end{align*}
	
		\begin{figure}
       \centering
			\includegraphics[width=8.5cm]{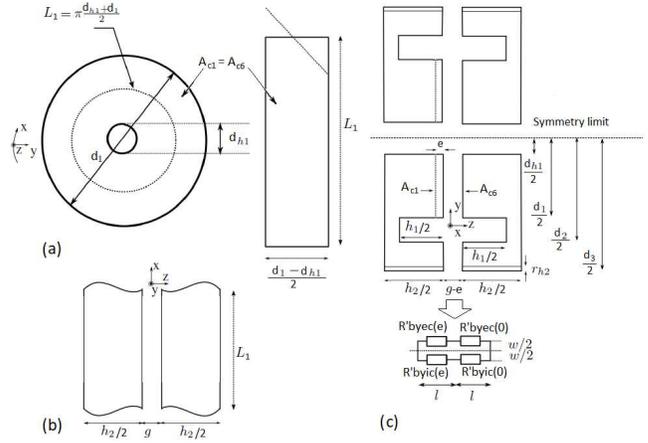}
	    \caption{Central gap in PM core. For $\mathcal{R}^{'}_{byic}$: $l=\frac{g}{2}$, $w=\frac{d_1-d_{h1}}{2}$ and $h=\frac{h_1}{2}-e$. For $\mathcal{R}^{'}_{byec}$: $l=\frac{g}{2}$, $w=\frac{d_1-d_{h1}}{2}$ and $h=\frac{h_2}{2}-e$.}
       \label{fig:PM_Core_intgap}
\end{figure}

   To obtain $\sigma_{ye}(x)$, the same reasoning used before for the \emph{x-z} plane is now applied to the \emph{y-z} plane, leading to
	\begin{align*}
			\mathcal{R}^{'}_{ye}(x)&=2\mathcal{R}^{'}_{byee}(x)//2\mathcal{R}^{'}_{byie}(x)=\frac{2\mathcal{R}^{'}_{byie}(x) \mathcal{R}^{'}_{byee}(x)}{\mathcal{R}^{'}_{byie}(x)+\mathcal{R}^{'}_{byee}(x)}\\
		\sigma_{ye}(x)&=\frac{\mathcal{R}^{'}_{ye}(x)}{\frac{g}{\mu_o w(x)}} 		
	\end{align*}
To take into account the overall effect of the non homogeneous gap width, an average fringing factor $\overline{\sigma_y}$ is obtained as
	\begin{align*}
		\overline{\sigma_{ye}}&=\frac{4}{(\pi-2\alpha)(d_2+d_3)}\int^{\frac{(\pi-2\alpha)(d_2+d_3)}{8}}_{-\frac{(\pi-2\alpha)(d_2+d_3)}{8}}{\sigma_{ye}(x)dx}  
\end{align*}
After some manipulation and finally applying symbolic integration tools,  $\overline{\sigma_{ye}}$ turns to be
		\begin{align*}
		\overline{\sigma_{ye}}&=\frac{a}{a+c}+\frac{8}{(\pi - 2\alpha)(d_3+d_2)} \vast\{ \frac{d_{h2}}{2}\left( 1-\frac{a}{a+c} \right) \\ 
			& +\frac{c}{2 \frac{r_{h2}}{d_{h2}}} \Biggl\{ \frac{\pi}{2}-\frac{a+c}{\sqrt{(a+c)^2-{r_{h2}}^2}} \Biggr[ \frac{\pi}{2} \\
			& +\mathrm{arctan} \left( \frac{r_{h2}}{\sqrt{(a+c)^2-{r_{h2}}^2}} \right) \Biggr] \Biggr\} \vast\}	\\
				a&=\frac{d_3-d_2}{2}	\\
				c&=\frac{2g}{\pi} \left( 1+\ln{\frac{\pi \sqrt{h_1 h_2}}{4g}} \right)
		\end{align*}

			The reluctance of an air gap lacking of fringing flux is given by
			\begin{align*}
			\mathcal{R}_{ggo}=\frac{g}{\mu_o A_{cg}}
			\end{align*}
			where $A_{cg}=A_{c5}$ is the core area in contact with the gap. 		
		The fringing factors $\sigma_{xe}$ and $\overline{\sigma_{ye}}$ take into account the fringing fluxes by introducing the effective air gap area $A_{gee}$ that virtually increases $A_{cg}$ in both $x$ and $y$ directions, thus reducing the reluctance. Consequently, the actual $\mathcal{R}_{gge}$ can be calculated as
			\begin{align}
			\label{eq:R_gge}
			\mathcal{R}_{gge}&=\sigma_{xe} \overline{\sigma_{ye}} \mathcal{R}_{ggo}=\frac{g}{\mu_o A_{gee}} \\
			A_{gee}&=\frac{A_{c5}}{\sigma_{xe} \overline{\sigma_{ye}}} \notag 
			\end{align}
				
\subsection{Central reluctance $\mathcal{R}_{ggc}$}

\begin{figure}
       \centering
			\includegraphics[width=8.5cm]{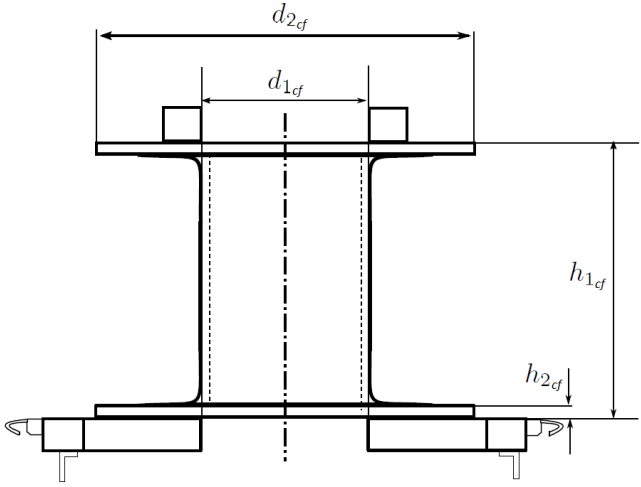}	
       \caption{Typical coil former for a PM core}
       \label{fig:PM_CoilFormer}
\end{figure}		
	The center gap reluctance $\mathcal{R}_{ggc}$ is obtained by considering the rounded central leg with its centered hole as an unfolded annulus as shown in Figure~\ref{fig:PM_Core_intgap}a, transforming $A_{c1}$ and $A_{c6}$ into equivalent rectangular surfaces. In the \emph{x-z} plane shown in Figure~\ref{fig:PM_Core_intgap}b, it is assumed that no fringing flux exists in the x-direction since there is an endless closed ring. On the contrary, in the \emph{y-z} plane shown in Figure~\ref{fig:PM_Core_intgap}c, there is one side in contact with the windings and the other side in contact with the small central hole. Note that the actual fringing flux in this side will be more constrained to flow than the existing over the opposite side. Thus in this case, the formulae used before cannot be applied in a straightforward way. As a workaround, we propose the use of a basic 2D reluctance per-unit-of-length that is a geometric average ($\mathcal{R}^{'}_{byec}$) between two extremes. On the one hand, the case with no central hole and thus no fringing flux ($\mathcal{R}^{'}_{byenf}$) and in the other hand, the case where the hole radius is large enough to impose no restrictions over the fringing flux to circulate ($\mathcal{R}^{'}_{byef}$), like at the other side of the central leg.  
	Consequently, in the x-direction
		\begin{align*}
	\sigma_{xc}=1
	\end{align*}
and in the y-direction we get 
				\begin{align}
		\mathcal{R}^{'}_{byef}(e)&=\frac{1}{\mu_o\left[\frac{d_1-d_{h1}}{2g}+ \frac{2}{\pi}\left(1 + \ln{ \frac{\pi \left(\frac{h_2}{2}-e \right)}{2g}} \right)\right]} \notag \\
		\mathcal{R}^{'}_{byenf}&=\frac{2g}{\mu_o(d_1-d_{h1})}	\notag\\
		\label{eq:Rbyec}
		\mathcal{R}^{'}_{byec}(e)&=\sqrt{\mathcal{R}^{'}_{byef}(e) \mathcal{R}^{'}_{byenf}}	\\
		\mathcal{R}^{'}_{byic}(e)&=\frac{1}{\mu_o\left[\frac{d_1-d_{h1}}{2g}+ \frac{2}{\pi}\left(1 + \ln{ \frac{\pi \left(\frac{h_1}{2}-e \right)}{2g}} \right)\right]} \notag \\
		\mathcal{R}^{'}_{byc}(e)&=\frac{\mathcal{R}^{'}_{byec}(e) \mathcal{R}^{'}_{byic}(e)}{\mathcal{R}^{'}_{byec}(e)+\mathcal{R}^{'}_{byic}(e)} \notag \\
		\mathcal{R}^{'}_{yc}&=\begin{Bmatrix} \mathcal{R}^{'}_{byc}(e=g)+\mathcal{R}^{'}_{byc}(e=0) & \mbox{if}& q_g=1 \\ 2 \mathcal{R}^{'}_{byc}(e=0)  & \mbox{if}& q_g=2 \notag
		\end{Bmatrix}\\		
		\sigma_{yc}&=\frac{\mathcal{R}^{'}_{yc}}{\frac{2g}{\mu_0(d_1-d_{h1})}} 		
			\end{align}
			Being now $A_{cg}=A_{c1}$, the actual $\mathcal{R}_{ggc}$ turns to be
		\begin{align}
		\label{eq:R_ggc}
		\mathcal{R}_{ggc}&=\sigma_{xc}\sigma_{yc}\mathcal{R}_{ggo}=\frac{g}{\mu_o A_{gec}} \\
		A_{gec}&=\frac{A_{c1}}{\sigma_{xc}\sigma_{yc}} \notag 			
			\end{align}	
	
	\section{Air gap length $g$} \label{sec:Length}
		
	According to \eqref{eq:R_gg_core}, to obtain $g$ as a function of $\mathcal{R}_{gg}$
	, the implicit equation
	\begin{align}
		\label{eq:Rggmin_solve}
			 \mathcal{R}_{gg}=\begin{Bmatrix} \mathcal{R}_{ggc}(g)	& \mbox{if}& q_g=1 \\	 \mathcal{R}_{ggc}(g)+\mathcal{R}_{gge}(g) & \mbox{if}& q_g=2
			\end{Bmatrix}
			\end{align}
	has to be numerically solved. As it is shown in \cite{3DAirgapReluctanceCalculations:Muhlethaler}, $\mathcal{R}_{gg}=0 \Leftrightarrow g=0$ and $\mathcal{R}_{gg}(g)$ is increasing with $g$. This makes easy to invert $\mathcal{R}_{gg}(g)$ by finding the solution of \eqref{eq:Rggmin_solve} within the closed interval $[g^*_{min}, g^*_{max}]$ defined next, employing a simple root-solver like Matlab's $fzero$ function.   
	An adequate $g^*_{min}$ would be
	\begin{align*}
			 g^*_{min}&=\frac{\mu_0 A_{cg} N^2}{L_i}	\\
			A_{cg}&=\begin{Bmatrix} A_{c1}  & \mbox{if}& q_g=1 \\ \frac{A_{c1}A_{c5}}{A_{c1}+A_{c5}}  & \mbox{if}& q_g=2 
		\end{Bmatrix}	
						\end{align*}
		which corresponds to the ideal case of no existing fringing flux and initial permeability $\mu_i \rightarrow \infty$; $L_i$ is the initial inductance and $N$ is the number of winding turns of the current inductor being designed.
		If $q_g=2$, $g^*_{max}$ should be selected as 
		\begin{align*}
		3 g^*_{min} \leq g^*_{max} \leq 10 g^*_{min}
		\end{align*}
		in most design cases, though there is not a definite limit because the spacers can be made as thick as desired. However, if $q_g=1$ then we are limited to $g^*_{max}=\frac{h_1}{2}$ which is the height of a core piece central leg. In this case, if the required $g$ were greater than $g^*_{max}$, then $q_g=2$ or a larger core should be initially chosen. 
\begin{figure*}
       \centering
			\includegraphics[width=18cm]{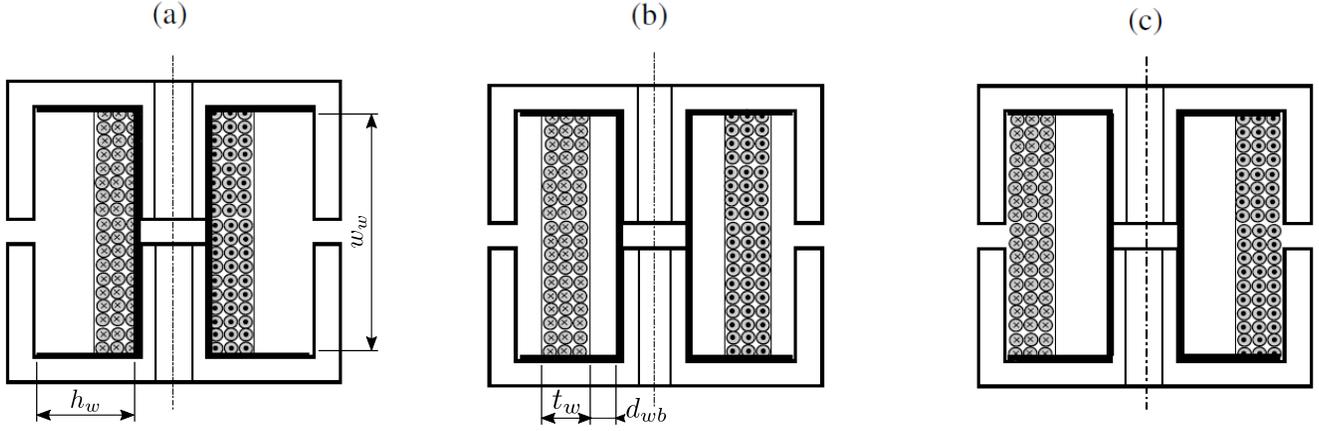}
       \caption{Alternative dispositions of the winding inside the coil former}
       \label{fig:Winding_inside_bobbin}
\end{figure*}	
		
		\section{Winding height, winding width and average turn length} \label{sec:CoilFormer}

	 The drawing of a typical coil former for PM cores is shown in Figure~\ref{fig:PM_CoilFormer}.
The minimum winding width $w_w$ and the maximum winding height $h_w$ are
\begin{align*}
w_w&=h_{1_{cf}}-2h_{2_{cf}}	\\
h_w&=\frac{d_{2_{cf}}-d_{1_{cf}}}{2}	
\end{align*}
The winding is arranged inside the coil former according to one of the alternatives depicted in Figure~\ref{fig:Winding_inside_bobbin}. The average length of a winding turn $l_t$ as a function of the winding thickness $t_w$ is then	
	\begin{align}
	\label{eq:lt}
		l_t \left( t_w \right)&=\pi \left( d_{1_{cf}}+2 ~ d_{wb}+t_w \right)
		\end{align}
		where the distance between the winding and the bobbin $d_{wb}$ is
		\begin{align*}
		d_{wb}&=\begin{Bmatrix} 0  & \mbox{inward alignment (cf. Figure~\ref{fig:Winding_inside_bobbin}.a)} \\ \frac{h_w-t_w}{2}  & \mbox{centered alignment (cf. Figure~\ref{fig:Winding_inside_bobbin}.b)} \\  h_w-t_w & \mbox{outward alignment (cf. Figure~\ref{fig:Winding_inside_bobbin}.c)} \end{Bmatrix} 
\end{align*}

\section{Conclusion} \label{sec:Conclusions}
This paper thoroughly develops the required core-specific parameters and functions exclusively related to the shape of Power Module (PM) cores, which are needed by %the design method presented in \cite{MyPaper:AVS_MR}
a comprehensive design-oriented inductor model. This is a fundamental tool to properly design and control any type of electronic power converter. The same procedures followed in this paper can be easily adapted to obtain the required data for other shapes of ferrite cores. Further work will be devoted in developing the exact expression of \eqref{eq:Rbyec}, the external part of the central gap basic reluctance in the y-z plane $\mathcal{R}^{'}_{byec}$, by extending the reasoning of \cite{3DAirgapReluctanceCalculations:Muhlethaler} to this particular structure. Experimental measurements of the initial inductance of several PM-core based inductors are in process, showing so far very good agreement with theoretical predictions.           

\section*{Acknowledgment}

The  first author wants to thank Dr. Hernan Haimovich for his guidance and constructive suggestions.

%\ifCLASSOPTIONcaptionsoff
  %\newpage
%\fi


\begin{thebibliography}{00}

%\bibitem{MyPaper:AVS_MR}
%A. Vazquez Sieber, M. Romero and H. Haimovich, \emph{A comprehensive method to determine the minimum number of turns and the optimum air gap length in ferrite-core based low-frequency-current biased AC filter inductors}, submitted to IEEE Trans. Power Electron. % 

\bibitem{TradeoffStudyHeatSinkOutputFilterVolumeGaNHEMTBasedSingle-PhaseInverter:Castellazzi}
E. Gurpinar and A. Castellazzi, \emph{Tradeoff Study of Heat Sink and Output Filter Volume in a GaN HEMT Based Single-Phase Inverter}, in IEEE Transactions on Power Electronics, vol. 33, no. 6, pp. 5226-5239, June 2018.

\bibitem{HybridActivePowerFilterGaNPowerStage5kWSinglePhaseInverter:Manchia}
R. Otero-De-Leon, L. Liu, S. Bala and G. Manchia, \emph{Hybrid active power filter with GaN power stage for 5kW single phase inverter}, 2018 IEEE Applied Power Electronics Conference and Exposition (APEC), San Antonio, TX, 2018, pp. 692-697

%\bibitem{OutputDv/DtFilterDesignCharacterization10kWSiCInverter:Mertens}
%J. Müller, T. Brinker, J. Friebe and A. Mertens, \emph{Output dv/dt Filter Design and Characterization for a 10 kW SiC Inverter}, IECON 2018 - 44th Annual Conference of the IEEE Industrial Electronics Society, Washington, DC, 2018, pp. 2122-2127.
%
%\bibitem{SmallSaturatingInductorsCompactSwitchingPowerSupplies:RinconMora}
%L. Milner and G. Rinc\'on-Mora, \emph{Small saturating inductors for more compact switching power supplies}, IEEJ Transactions on Electrical and Electronic Engineering, vol. 7, no. 1, pp. 1-5, 2011. 

\bibitem{200COperationDC-DCConverterSiCPowerDevices:Jordan}
B. Ray, H. Kosai, J. D. Scofield and B. Jordan, \emph{200°C Operation of a DC-DC Converter with SiC Power Devices}, APEC 07 - Twenty-Second Annual IEEE Applied Power Electronics Conference and Exposition, Anaheim, CA, USA, 2007, pp. 998-1002.

\bibitem{99EfficientThree-phaseBuck-typeSiCMOSFETPFCRectifierMinimizingLifeCycleCostDCdataCenters:Kolar}
L. Schrittwieser, J. W. Kolar and T. B. Soeiro, \emph{99\% Efficient three-phase buck-type SiC MOSFET PFC rectifier minimizing life cycle cost in DC data centers}, CPSS Transactions on Power Electronics and Applications, vol. 2, no. 1, pp. 47-58, 2017.

\bibitem{High-VoltageSiC-BasedBoostPFCLEDApplications:Salamero}
A. Leon-Masich, H. Valderrama-Blavi, J. M. Bosque-Moncusí and L. Martínez-Salamero, \emph{A High-Voltage SiC-Based Boost PFC for LED Applications}, IEEE Transactions on Power Electronics, vol. 31, no. 2, pp. 1633-1642, Feb. 2016.

\bibitem{DesignLossAnalysisHighFrequencyPFCconverter:Song}
Q. Li, H. Y. Zhao and J. C. Song, \emph{Design and loss analysis of the high frequency PFC converter}, 2015 IEEE International Conference on Applied Superconductivity and Electromagnetic Devices (ASEMD), Shanghai, 2015, pp. 124-125.

\bibitem{TransformerInductorDesign:McLyman}
C.~Wm.~T. McLyman, \emph{Transformer and Inductor Design Handbook}, 3rd~ed.\hskip 1em plus
  0.5em minus 0.4em\relax New York, USA: Marcel-Dekker, 2004.

\bibitem{High-FrequencyMagneticComponents:Kazimierczuk}
M.~K. Kazimierczuk, \emph{High-Frequency Magnetic Components}, 2nd~ed.\hskip 1em plus
  0.5em minus 0.4em\relax West Sussex, England: John Wiley \& Sons, 2014.	

\bibitem{PowerElectronics:Mohan}
N. Mohan, T.~M. Undeland and W.~P. Robbins, \emph{Power Electronics. Converters, Applications, and Design}, 3rd~ed.\hskip 1em plus
  0.5em minus 0.4em\relax New York, USA: John Wiley \& Sons, 2003.

\bibitem{InductorsTransformersPowerElectronics:Bossche}
A. Van den Bossche and V.~C. Valchev, \emph{Inductors and Transformers for Power Electronics}, 1st~ed.\hskip 1em plus
  0.5em minus 0.4em\relax Boca Raton, USA: Taylor \& Francis, 2005.
	
\bibitem{CalculationEffectiveParametersMagneticPieceParts:IEC60205}
 \emph{Calculation of the effective parameters of magnetic piece parts}, IEC 60205 edition 3.1, International Electrotechnical Commission, \hskip 1em plus
  0.5em minus 0.4em\relax Geneva, Switzerland, Aug. 2009.

\bibitem{PM-cores_made_of_magnetic_oxides_and_associated_parts:IEC61247}
\emph{PM-cores made of magnetic oxides and associated parts - Dimensions}, IEC 61247 edition 1.0b., International Electrotechnical Commission, \hskip 1em plus
  0.5em minus 0.4em\relax Geneva, Switzerland, 1995.
	
	\bibitem{FerritePM62/49:TDK-EPCOS}
EPCOS AG, \emph{PM 62/49. Core and accessories}, Ferrite and Accessories, May 2017.
 [Online]. Available://www.tdk-electronics.tdk.com/inf/80/db/fer/pm\_62\_49.pdf.

\bibitem{3DAirgapReluctanceCalculations:Muhlethaler}
J. M\"{u}hlethaler, J.~W. Kolar and A. Ecklebe, \emph{A Novel Approach for 3D Air Gap Reluctance
Calculations}, in Proc. of the International Conference on Power Electronics - ECCE Asia, pp. 446-452, Jun. 2011.


\end{thebibliography}
\end{document}